\documentclass[reprint,amsmath,amssymb,aps,superscriptaddress]{revtex4-2}
\usepackage{graphicx}
\usepackage{dcolumn}
\usepackage{hyperref}
\usepackage[mathlines]{lineno}
\usepackage{bbold}
\usepackage{tabularx}
\usepackage{cleveref}
\usepackage{physics}
\usepackage{siunitx}
\usepackage{multirow}
\usepackage{subcaption}
\usepackage{caption}
\usepackage{booktabs}
\usepackage{bm}
\usepackage{lipsum} 

\draft 

\begin{document}


\title{Quantum Spectral Clustering: Comparing Parameterized and Neuromorphic Quantum Kernels} 

\author{Donovan Slabbert}
\email{donovanslab@mweb.co.za}
\affiliation{Department of Physics, Stellenbosch University, Stellenbosch, 7600, South Africa}
\affiliation{National Institute for Theoretical and Computational Sciences (NITheCS), Stellenbosch, 7600, South Africa}
\author{Dean Brand}
\email{dean.brand@nithecs.ac.za}
\affiliation{Department of Physics, Stellenbosch University, Stellenbosch, 7600, South Africa}
\affiliation{National Institute for Theoretical and Computational Sciences (NITheCS), Stellenbosch, 7600, South Africa}
\author{Francesco Petruccione}
\email{petruccione@sun.ac.za}
\affiliation{School of Data Science and Computational Thinking, Stellenbosch University, Stellenbosch, 7600, South Africa}
\affiliation{Department of Physics, Stellenbosch University, Stellenbosch, 7600, South Africa}
\affiliation{National Institute for Theoretical and Computational Sciences (NITheCS), Stellenbosch, 7600, South Africa}

\date{\today}

\begin{abstract}
We compare a parameterized quantum kernel (pQK) with a quantum leaky integrate-and-fire (QLIF) neuromorphic computing approach that employs either the Victor-Purpura or van Rossum kernel in a spectral clustering task, as well as the classical radial basis function (RBF) kernel. Performance evaluation includes label-based classification and clustering metrics, as well as optimal number of clusters predictions for each dataset based on an elbow-like curve as is typically used in $K$-means clustering. The pQK encodes feature vectors through angle encoding with rotation angles scaled parametrically. Parameters are optimized through grid search to maximize kernel-target alignment, producing a kernel that reflects distances in the feature space. The quantum neuromorphic approach uses population coding to transform data into spike trains, which are then processed using temporal distance metrics. Kernel matrices are used as input into a classical spectral clustering pipeline prior to performance evaluation. For the synthetic datasets and \texttt{Iris}, the QLIF kernel typically achieves better classification and clustering performance than pQK. However, on higher-dimensional datasets, such as a preprocessed version of the Sloan Digital Sky Survey (\texttt{SDSS}), pQK performed better, indicating a possible advantage in higher-dimensional regimes.
\end{abstract}

\pacs{}

\maketitle 

\section{Introduction}
\label{sec:Introduction}

Efforts to achieve quantum advantage and demonstrate practical quantum utility continue to drive research in the field of quantum computing \cite{preskill2012quantum, arute2019quantum, kim2023evidence}. This is especially true for quantum machine learning, where the supposed computational power of quantum computers is combined with machine learning \cite{Biamonte2017, schuld2015introduction, schuld2021machine}. The reality of the current state of quantum computing demands a more realistic outlook. Noisy intermediate-scale quantum (NISQ) devices \cite{preskill2018quantum, bharti2022noisy} face many challenges, some of which include decoherence \cite{shor1995scheme, brandt1999qubit}, qubit connectivity \cite{linke2017experimental}, and error correction \cite{steane1996error, lidar2013quantum}. Despite these challenges, the application of machine learning through quantum devices, even if only simulated, has been widely investigated \cite{cerezo2022challenges}.

A common approach to quantum machine learning is to use a hybrid quantum and classical approach \cite{liu2021hybrid, benedetti2019parameterized, slabbert2025classical}. One such approach is to use quantum kernels \cite{huang2021power, liu2021rigorous, schuld2019quantum, havlivcek2019supervised} in a quantum support vector machine (QSVM) \cite{MariaSchuld2021, schuld2021supervised}, which fits a classical SVM using a quantum kernel. This approach can be thought of as a quantum-enhanced SVM, if some advantage for using a quantum kernel can be found. QSVMs have been studied and applied extensively and fall under the category of supervised machine learning. Unsupervised quantum machine learning \cite{otterbach2017unsupervisedmachinelearninghybrid, lloyd2013quantumalgorithmssupervisedunsupervised} has been considered and implemented before, but is severely underrepresented in the current literature.

Spectral clustering \cite{ng2001spectral, von2007tutorial} is a well-known approach to unsupervised machine learning for clustering that also requires a kernel, which can be quantum. This implies that we can implement quantum-enhanced spectral clustering by simply feeding a quantum kernel to already established spectral clustering tools. Quantum clustering \cite{poggiali2022clustering, poggiali2024quantum, kavitha2023quantum, khan2019k}, and in particular quantum spectral clustering \cite{kerenidis2021quantum, li2022quantum}, have been previously explored. However, like most kernel-based methods, these approaches suffer from quadratic scaling with respect to the number of samples, since the kernel must be evaluated for every pairwise combination in the dataset. The more pertinent issue, if we ignore the scaling for now, is that the quantum kernel is not always a good representation of similarity or distance between two samples. This is because classical information resides in feature space, but once encoded in a quantum computer, it is mapped to a high-dimensional Hilbert space \cite{PhysRevA.103.032430}. In this setting, conventional notions of distance often become unreliable, making traditional clustering approaches such as $K$-means that use Euclidean distance or other classical distance-based kernel functions unsuitable. As a result, some researchers have explored quantum-native clustering methods that operate directly on quantum states using gates \cite{NEURIPS2019_16026d60}. Another approach is to focus on designing classical kernels that account for quantum encoding in a meaningful way \cite{hubregtsen2022training}. Spectral clustering remains viable in this context, provided that a representative kernel can be identified and utilized that captures the relevant structure in the Hilbert space.

The goal of this paper is to introduce and compare two quantum kernel approaches, from two distinct perspectives, to find such a representative kernel. Parameterization of a quantum kernel and finding the optimal parameters in a trainable quantum kernel \cite{alvarez2025benchmarking, rodriguez2024satellite}. We choose to do this by parameterizing the encoding, making it possible to find the optimal encoding for the given model architecture for the task under consideration. The encoding strategy is often one of the most important factors in determining the expressibility of a quantum model \cite{PhysRevA.103.032430}. The other approach is through quantum neuromorphic computing \cite{brand_quantum_2024,markovic_quantum_2020,ghosh_quantum_2021}, where samples are converted into binary spike trains \cite{eshraghian_training_2023} and fed into kernels based on distance metrics to quantify the similarity of the generated spike trains. The above are the two approaches that will be directly compared and contrasted.

One real-world use case that we focus on is astronomy. Astronomical surveys routinely produce large amounts of unlabeled data, which require unsupervised methods to identify meaningful structures among celestial objects such as stars, galaxies, and quasars in datasets such as the Sloan Digital Sky Survey (\texttt{SDSS}) \cite{york2000sloan, ball2010data}. However, both traditional and quantum-enhanced clustering methods face challenges when the kernel or feature representation does not align well with the underlying data distribution \cite{ball2010data, smola1998learning, schuld2015introduction, abbas2021power}. This paper investigates whether more expressive quantum kernels, either through trainable quantum circuits that learn optimal encodings or through biologically inspired quantum neuromorphic kernels that operate on spike-based representations, can offer improved clustering performance. By comparing these two approaches within the same spectral clustering framework, we aim to assess their utility as quantum kernel methods for clustering, with the primary goal of determining which kernel achieves superior performance based on clustering and classification metrics.

The rest of the paper is structured as follows: Section \ref{sec:Datasets} briefly introduces the datasets used. Section \ref{sec:Theory} explains the spectral clustering, trainable gate-based quantum kernels, and quantum neuromorphic computing kernels. Section \ref{sec:Results} discusses the results of applying spectral clustering using our kernels, and Section \ref{sec:Conclusion} concludes the paper.

\section{Datasets}
\label{sec:Datasets}

This section discusses all datasets used in our clustering pipeline. Please note that only 300 samples per dataset were used for evaluation. For pQK, both training and testing were performed on 300 samples.

\subsection{Synthetic Datasets}

We include commonly used synthetic datasets for benchmarking purposes during clustering. These are \texttt{Blobs}, \texttt{Circles}, and \texttt{Moons}, all imported through \texttt{scikit-learn} \cite{pedregosa2011scikit}. Each sample has two features, the x- and y-coordinate, and we take the number of clusters to represent different classes for the classification part. We manually assign a label to each cluster. \texttt{Blobs} will have three classes, and \texttt{Moons} and \texttt{Circles} will only have two. The \texttt{Blobs}, \texttt{Moons}, and \texttt{Circles} datasets were generated with the following parameters, respectively: a cluster standard deviation of 1.40 for \texttt{Blobs}, noise of 0.075 for \texttt{Moons}, and noise of 0.1 with a factor of 0.2 for \texttt{Circles}.

\subsection{\texttt{Iris} and \texttt{SDSS}}

\texttt{Iris} \cite{iris_53} and \texttt{SDSS} \cite{a_o_clarke_2020_3768398, clarke2020identifying} are included for a real-world application. \texttt{Iris} samples have four features per sample, and \texttt{SDSS} has ten features. Please note that the \texttt{SDSS} dataset used, is not the original Sloan Digital Sky Survey, but instead a derivative that was preprocessed and curated from \texttt{SDSS}. It has also been feature selected to include only the ten most important features. Both \texttt{Iris} and \texttt{SDSS} have a ground truth, which means that we have class labels. Both datasets include three classes or clusters. We also include a balanced version of \texttt{SDSS}, balanced through sampling, which we refer to as \texttt{SDSS}\textsuperscript{*}.

\section{Theory and Implementation}
\label{sec:Theory}

\subsection{Spectral Clustering}
\label{sec:Spectral_Clustering}

Spectral clustering \cite{ng2001spectral, von2007tutorial} is a graph-theoretic approach to clustering that excels at discovering complex, non-convex cluster structures in data. At its core, the method transforms data into a graph-based representation, where points are treated as nodes and edges encode similarity between them. The algorithm leverages the spectrum (i.e. eigenvalues and eigenvectors) of a matrix derived from the data (most notably the Laplacian of a similarity graph) to partition the data into meaningful groups. This approach is particularly effective in cases where traditional algorithms like $K$-means fail due to assumptions of linear separability or isotropic cluster shapes \cite{von2007tutorial}.

The process begins with the construction of a kernel matrix, also referred to as the similarity matrix or affinity matrix. Given a dataset $\{\vec{x_1}, \vec{x_2}, \dots, \vec{x_n}\} \subset \mathbb{R}^d$, one common choice classically is the Gaussian (or RBF) kernel \cite{smola1998learning}:

\[
K_{ij} = \exp\left(-\frac{\|\vec{x_i} - \vec{x_j}\|^2}{2\sigma^2}\right) = \exp\left(-\gamma \|\vec{x_i} - \vec{x_j}\|^2\right),
\]

\noindent where $\sigma$ ($\gamma = \frac{1}{2\sigma^2}$) is a hyperparameter controlling the width of the neighbourhood in which points are considered similar. The resulting matrix $K \in \mathbb{R}^{n \times n}$ is symmetric and positive semi-definite, encoding pairwise similarities between data points. Importantly, the use of a kernel function allows the algorithm to operate in an implicit high-dimensional feature space without explicitly computing the transformation, which is an application of the kernel trick \cite{scholkopf1999advances, boser1992training} familiar from kernelized support vector machines.

Once the kernel matrix is formed, a graph Laplacian is constructed to analyze the connectivity structure of the graph. There are several variations, but a commonly used form is normalized Laplacians:

\begin{equation}
L_{\text{sym}} = I - D^{-1/2} K D^{-1/2}, \quad L_{\text{rw}} = I - D^{-1}K
\end{equation}

\noindent where $D$ is the degree matrix defined as $D_{ii} = \sum_j K_{ij}$. These Laplacians encode local smoothness and global structure of the graph. The normalized Laplacians are often preferred for their numerical stability and scale invariance and is used by default in \texttt{scikit-learn}. Spectral clustering then proceeds by computing the first $k$ eigenvectors $\{\vec{u_1}, \dots, \vec{u_k}\}$ of the chosen Laplacian, specifically those corresponding to the smallest eigenvalues (excluding zero if present). These eigenvectors are stacked as columns in a matrix $U \in \mathbb{R}^{n \times k}$, and each row of $U$ is treated as a new feature representation of a data point.

In this transformed space, the final step is to apply a conventional clustering algorithm (typically $K$-means) to the rows of $U$. The intuition is that the eigenvectors reveal a low-dimensional embedding in which clusters are more linearly separable. From a mathematical perspective, this approach can be interpreted as a relaxation of the normalized cut problem \cite{shi2000normalized}, which seeks to partition a graph into disjoint sets while minimizing the total edge weight between the sets relative to their volume. Spectral clustering thus offers a principled and elegant framework for grouping data, grounded in spectral graph theory and enabled by the kernel matrix’s ability to model intricate relationships. More importantly it allows its use in hybrid quantum-classical machine learning pipelines, where the spectral clustering is fitted classically using a quantum kernel. Note that all spectral clustering implementations in this work were carried out using the built-in functionality of \texttt{scikit-learn}.

\subsection{Trainable Quantum Kernel}
\label{sec:Trainable_Quantum_Kernel}

\begin{figure*}[t!]
    \centering
    \includegraphics[width=\textwidth]{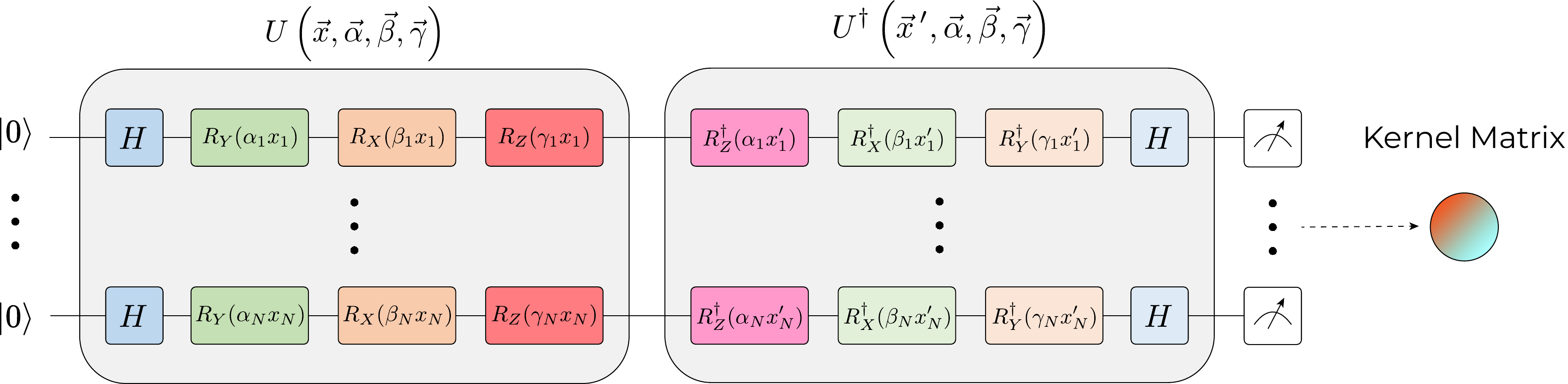}
    \captionsetup{justification=raggedright}
    \caption{Architecture of the trainable parameterized encoding quantum kernel. A full kernel matrix is found for each pair of samples via a measurement using a projector matrix. The kernel will be used during spectral clustering.}
    \label{fig:Q_trainable}
\end{figure*}

\subsubsection{Parameterized Encoding}
\label{sec:Parameterized Quantum Kernel}

Quantum kernels refer to kernel functions evaluated on a quantum computer, where the feature space is the quantum mechanical Hilbert space in which the quantum state exists after data encoding, typically defined by a quantum circuit known as a feature map or state preparation circuit \cite{schuld2019quantum, schuld2020circuit}. These quantum-enhanced kernels are particularly useful in hybrid classical-quantum machine learning models such as the support vector machine (SVM) \cite{cortes1995support}, allowing non-linear separation of data in a high-dimensional space that may be challenging to access classically.

In our implementation, we employ a quantum fidelity kernel \cite{schuld2019quantum, havlivcek2019supervised}, where fidelity in this case is a similarity measure based on the overlap between quantum states corresponding to classical inputs $\vec{x}$ and $\vec{x}'$. Each data point or feature vector $\vec{x}$ is mapped to a quantum state $|\phi(\vec{x})\rangle$ via a unitary state preparation block $U(\vec{x})$. Similarly, a second feature vector $\vec{x}'$ is mapped to another quantum state $|\phi(\vec{x'})\rangle$ through a unitary $U^{\dagger}(\vec{x'})$, where the dagger represents the Hermitian conjugate of the unitary. This ensures the encoding of both feature vectors, $\vec{x}$ and $\vec{x}'$, in their respective quantum states. Please note that all features are scaled and normalized to $[0, \pi]$ before encoding.

Instead of relying on SWAP tests or ancilla-assisted measurements \cite{schuld2017implementing}, which are more costly in terms of number of qubits and circuit depth, we directly compute the kernel via a projector-based measurement in the zero state. The quantum kernel function is expressed as:

\begin{equation}
    K(\vec{x},\vec{x}') = |\bra{0\ldots0} U^{\dagger}(\vec{x}') U(\vec{x}) \ket{0\ldots0}|^2,
\end{equation}

\noindent where the measurement is with respect to the rank-one projector:

\begin{equation}
    \rho = |00\ldots0\rangle \langle 00\ldots0|.
    \label{eq=pmatrix}
\end{equation}

\noindent In summary this is equivalent to applying $U(\vec{x})$ to the initial state, followed by the inverse feature map $U^\dagger(\vec{x}')$, and then projecting back onto the $\ket{0}^{\otimes n}$ reference state. This is essentially calculating the fidelity between the two states $|\phi(\vec{x})\rangle$ and $|\phi(\vec{x}')\rangle$, since:

\begin{equation}
    K(\vec{x}, \vec{x}') = |\langle \phi(\vec{x}') | \phi(\vec{x}) \rangle|^2 = \text{Fidelity}^2.
\end{equation}

\noindent To construct the state \( |\phi(\vec{x})\rangle \), we use a simple and shallow parameterized quantum circuit that encodes classical features using quantum gates. Specifically, we apply three rotation gates per qubit: a rotation around the X-axis, \( R_x(x_i) \), followed by a rotation around the Y-axis, \( R_y(x_i) \), and a rotation around the Z-axis, \( R_z(x_i) \), where \( x_i \) represents the \(i\)-th component of the classical feature vector \( \vec{x} \). These rotations are defined as:

\begin{align}
    R_x(x_i) &= \exp\left(-i \frac{x_i}{2} X\right), \\
    R_y(x_i) &= \exp\left(-i \frac{x_i}{2} Y\right), \\
    R_z(x_i) &= \exp\left(-i \frac{x_i}{2} Z\right),
\end{align}

\noindent where \( X \), \( Y \), and \( Z \) are Pauli operators. For a classical input vector \( \vec{x} = (x_1, x_2, \ldots, x_n) \), each feature is mapped to rotation angles directly by using \( x_i \) as the corresponding angle for each rotation gate.

The parameter sets $\alpha$, $\beta$, and $\gamma$ are introduced as scaling parameters that scale the rotations applied to each qubit. These parameters can be fixed or trainable, depending on the specific model configuration. Each feature value in the classical input vector $\vec{x} = (x_1, x_2, \ldots, x_n)$ is encoded using all three rotation gates. Specifically, for each qubit $i$, the first rotation is scaled by the parameter $\alpha_i$, the second by $\beta_i$, and the third by $\gamma_i$. These parameters allow for more flexible and expressive encodings of the feature space, as each rotation can be independently scaled to achieve a desired transformation.

Since three rotation gates are applied per qubit, we require $3n$ parameters to define the unitary $U(\vec{x})$ for $n$ qubits. The entire feature map includes both $U(\vec{x})$ and its adjoint $U^\dagger(\vec{x}')$, each applying the same sequence of parameterized rotations. However, to ensure that samples are encoded in a consistent and comparable manner, and to avoid introducing asymmetries between inputs, we reuse the same parameters for both parts of the transformation. This design not only preserves symmetry between the encoded inputs $\vec{x}$ and $\vec{x}'$, but also improves computational efficiency by reducing the total number of trainable parameters to just $3n$. This compact yet expressive encoding enables the quantum kernel to better capture complex relationships in the input space through a meaningful similarity measure. The rotation gates are parameterized by:

\begin{align}
    R_x'(x_i) = R_x(\alpha_j x_i), \\
    R_y'(x_i) = R_y(\beta_j x_i), \\
    R_z'(x_i) = R_z(\gamma_j x_i),
\end{align}

\noindent where $\alpha_i$, $\beta_i$, and $\gamma_i$ are the scaling parameters (either fixed or trainable).

This parameterized rotation-based encoding strategy is efficient in both depth and hardware execution. Since each qubit undergoes only three rotations per unitary, the circuit depth remains relatively shallow, which reduces sensitivity to noise and decoherence \cite{preskill2018quantum}. This is a crucial factor for current noisy intermediate-scale quantum (NISQ) devices. Furthermore, since we do not require ancillary qubits or entangling gates to compute the kernel, the method scales more favourably and allows for straightforward implementation on real quantum hardware or simulators.

In summary, our approach allows for efficient and expressive quantum kernel evaluations by combining a shallow parameterized circuit (for state encoding) with a measurement strategy that leverages the fidelity or overlap between quantum states. This makes it well-suited to kernel-based quantum machine learning models such as the quantum-enhanced SVM, especially when operating under the constraints of limited quantum resources, but more importantly allows the use of a trainable quantum kernel for spectral clustering purposes by finding the optimal encoding through rotation gates. The final measured value is the fidelity, which represents the similarity between two quantum states and will be used in the kernel target alignment (KTA) loss function explained in the next section.

The quantum circuit can be seen in Figure \ref{fig:Q_trainable}

\subsubsection{Kernel Target Alignment (KTA)}
\label{sec:KTA}

In order to find a quantum kernel that is representative of distance in the feature-embedded Hilbert space, it should be trained in some way before being used for spectral clustering purposes. Since we have access to ground truth labels, we can utilize the labels during training. We choose the kernel target alignment (KTA) \cite{wang2015overview, cristianini2001kernel} as the metric to define how well a kernel represents distance in a given feature space. The KTA is defined as

\begin{equation}
\text{KTA}(K, Y) = \frac{\mathrm{Tr}(K^T YY^T)}{\sqrt{\mathrm{Tr}(K^T K)} \cdot \sqrt{\mathrm{Tr}((YY^T)^2)}},
\end{equation}

\noindent where $K$ is the kernel matrix and $Y$ is a matrix of one-hot encoded \cite{bishop2006pattern} class labels. In one-hot encoding, each label is represented as a binary vector of length equal to the number of distinct classes. For a dataset with $C$ classes and $n$ samples, $Y \in \mathbb{R}^{n \times C}$ is constructed such that each row corresponds to a sample and contains a 1 in the column of its class, and 0s elsewhere. This formulation allows the label structure to be embedded directly into the alignment computation with the kernel matrix.

We can use the KTA to construct a loss function:

\begin{equation}
L(K,Y) = 1 - \text{KTA}(K,Y),
\end{equation}

\noindent such that minimizing $L(K,Y)$ is equivalent to maximizing the alignment between the quantum kernel and the label-informed similarity matrix. A maximized KTA implies that the kernel captures class-based structure in the Hilbert space as accurately as possible.

While gradient-based optimization is a natural approach to training parameters in many machine learning models, quantum circuits are known to suffer from barren plateaus \cite{mcclean2018barren}, which are regions in parameter space where the gradient vanishes exponentially with the number of qubits. This makes training deep or even moderately parameterized quantum circuits especially difficult. To avoid this issue, we opted for a grid search strategy over the kernel parameters.

In our grid search, the parameters were randomly initialized in the range $[0, 2]$, and the classical input features were scaled to lie within $[0, \pi]$. This combination ensures that the rotations $R_x(\alpha_j x_i)$, $R_y(\beta_j x_i)$, and $R_z(\gamma_j x_i)$ sweep through their entire respective directions on the Bloch sphere. Consequently, the encoded states are able to fully explore the representational capacity of the circuit in each rotational axis, without requiring gradient-based optimization. Note that two rotational directions would be sufficient to span the Bloch sphere \cite{nielsen2010quantum}; however, a third rotation is included to enhance the expressivity of the model.

When labels are unavailable, the goal of training a kernel is to find a meaningful similarity or distance structure in the data that facilitates clustering, without explicit reference to labels. Common objectives include \cite{von2007tutorial, ng2001spectral, wu2009choosing}, but are not limited to: Maximizing the spectral gap of the kernel matrix, which helps ensure clear separation between clusters in the feature space, improving the likelihood of discovering distinct groupings. Maximizing cluster coherence encourages the kernel to pull together samples within the same cluster, ensuring that the learned representation reflects tight, well-defined clusters. Maximizing inter-cluster distance (by minimizing fidelity) ensures that different clusters are pushed apart in the feature space, increasing the separability of the learned clusters. Training, in this context, refers to optimizing the kernel parameters to maximize the spectral gap, cluster coherence, and inter-cluster distance, focusing on the kernel matrix that reveals meaningful cluster structures without relying on labels. However, since all datasets considered include labels, there is no reason not to leverage them when available.

\subsection{Quantum Neuromorphic Neuron and Kernels}
\label{sec:Quantum_Neuromorphic_Neuron_and_Kernels}

Quantum neuromorphic computing has recently emerged as a promising and resource-efficient alternative for processing classical data on current NISQ-era quantum hardware. A recent example of this was in the construction of a quantum leaky integrate-and-fire (QLIF) neuron which was used in the composition of quantum spiking neural networks (QSNN) for image classification tasks \cite{brand_quantum_2024}. The proposed QLIF neuron is not, however, limited strictly to neural networks, but is a versatile computational unit, capable of processing data with a temporal dimension on a quantum computer with high resource efficiency.

We propose the idea of using quantum spiking neurons for unsupervised learning in the form of clustering. This makes use of a series of data processing concepts, which are threaded together for the first time in this work. This section will introduce and describe the functionality of each of these concepts and how they are connected to be suited for spectral clustering.

\subsubsection{Leaky Integrate-and-Fire Neuron}
\label{sec:Leaky_Integrate_and_Fire_Neuron}

\begin{figure*}[t!]
    \centering
    \includegraphics[width=\textwidth]{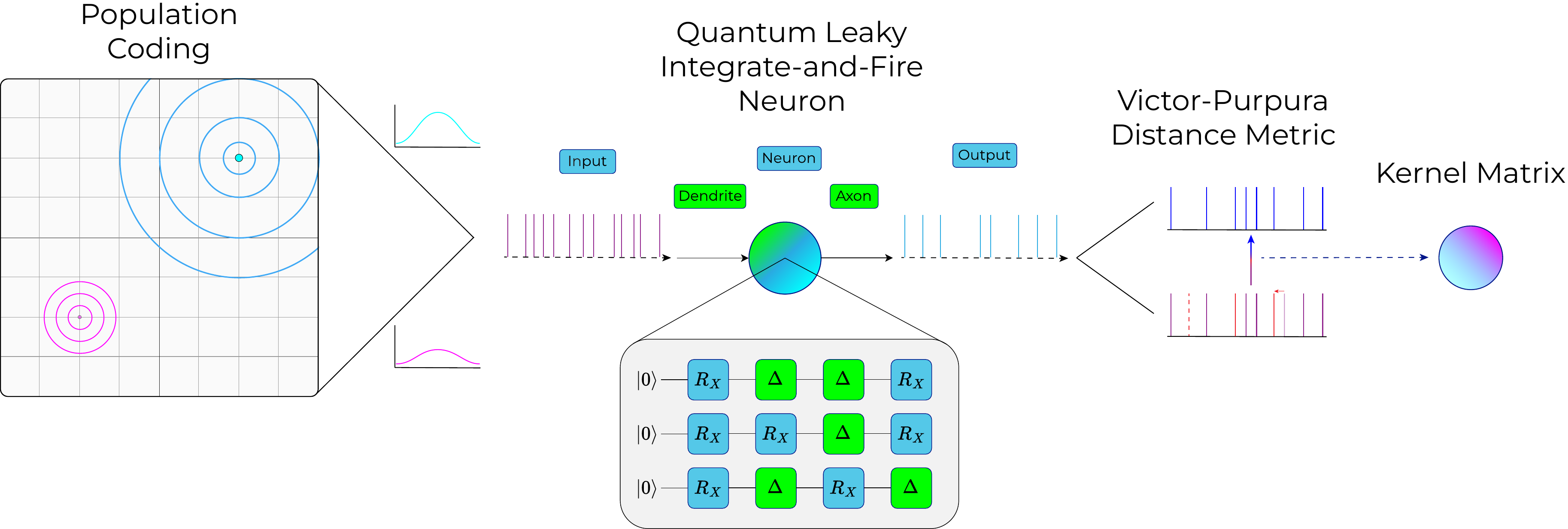}
    \captionsetup{justification=raggedright}
    \caption{Architecture of the quantum neuromorphic clustering algorithm. Coordinate data is encoded as spike trains through a population coding scheme, before processing the spike trains through the QLIF neuron. This produces output spikes that are all compared using the Victor-Purpura (or van Rossum) distance metric. These distances are used to compose a kernel matrix for spectral clustering.}
    \label{fig:QLIF_SC}
\end{figure*}

In the realm of neuromorphic computing, there are a myriad of options for computational units, which extend far beyond the typical classical neuron used in many machine learning tasks. Neuromorphic spiking neurons use binary activation pulses of current that encode temporal information as well as input intensity. Classical neurons, however, represent only mean spike firing rates, compressing the temporal dimension and losing information in the process \cite{haykin_neural_2009}. Spiking models offer significantly more information density at a much greater efficiency than classical artificial neurons on modern computer hardware.

Within the collection of spiking neuron models, there are several which offer varying levels and trade-offs of complexity, efficiency, and biological plausibility. A sweet-spot choice is that of the leaky integrate-and-fire (LIF) neuron, first studied in 1907 using biological systems \cite{brunel_lapicques_2007, burkitt_review_2006}. This neuron is a popular choice for highly performant machine learning tasks, due to its balance of simplicity and efficiency while being able to handle temporal data better than its classical counterpart \cite{yamazaki_spiking_2022,wang_supervised_2020}.

The behavior of the biological LIF neuron can be represented as a resistor-capacitor (RC) circuit acting as a low-pass filter \cite{hodgkin_quantitative_1952,eshraghian_training_2023}. In this model, the membrane potential, $U(t)$, accumulates due to input current spikes and returns to a stable state in the absence of input. This is described by the equation
\begin{equation}
\label{eq:rc_differential}
    \tau\dv{U(t)}{t} = -U(t) + I_\mathrm{in}R,
\end{equation}
with the time constant $\tau = RC$. By solving the differential equation and using the forward Euler method for approximation, we derive a time-discretized formula for the membrane potential
\begin{equation}
\label{eq:lif_discrete}
    U[t] = \beta U[t-1] + (1-\beta)I_\mathrm{in}[t],
\end{equation}
where the decay rate is denoted by $\beta = \exp(-1/\tau)$. The `fire' mechanism of the LIF neuron is triggered when the membrane potential surpasses the threshold value, $U_\mathrm{thr}$, leading the neuron to release an output spike by discharging the accumulated membrane potential.

To adapt the neuron expression for deep learning applications, the input current term can be interpreted as a weighted sum of a binary input spike, $X[t]$, and expressed in a simplified form as $I_\mathrm{in}[t] = WX[t]$, where $W$ is a learnable weight. This simplification, combined with the fire-reset mechanism, results in the expression
\begin{equation}
\label{eq:lif_final}
    U[t] = \underbrace{\beta U[t-1]}_\mathrm{decay} + \underbrace{WX[t]}_\mathrm{input} - \underbrace{S_\mathrm{out}[t-1]U_\mathrm{thr}}_\mathrm{reset},
\end{equation}
where the output spike is generated under the condition
\begin{equation}
\label{eq:lif_fire_condition}
    S_\mathrm{out}[t] = \begin{cases}
            1,\quad\text{if } U[t]>U_\mathrm{thr} \\
            0,\quad\text{otherwise}.
    \end{cases}
\end{equation}

\subsubsection{Quantum Leaky Integrate-and-Fire Neuron}
\label{sec:Quantum_Leaky_Integrate_and_Fire_Neuron}

To mimic the dynamics of the LIF neuron on a quantum computer, we follow the proposal and derivation of the original QLIF by Brand and Petruccione \cite{brand_quantum_2024}. In short, the aspects of the LIF neuron that need to be implemented are the 3 basic ingredients: spiking, decaying, and thresholding.

To encode an input stimulus within a quantum circuit, we employ a rotation gate to induce a spike by adjusting the angle according to the intensity of the input $\theta$. This action transitions the qubit from its initial state, $\ket{0}$, to an excited state $R_X(\theta)\ket{0}$, resulting in an immediate spike in the excited state population, similar to the membrane potential in classical models. Although the threshold mechanism operates similarly in both models, in the quantum approach, it is constrained to the range from $0$ to $1$. Regarding the exponential decay observed without an input spike, the quantum model excels by leveraging environmental noise. In the scenario of an open quantum system involving a quantum circuit, an excited state undergoing temporal evolution experiences $T_1$ relaxation, which leads to decay back to the ground state \cite{brand_markovian_2024}. The $T_1$ relaxation can be closely represented by an exponential decay function, precisely aligning with the neuron leak requirement.

When an input spike occurs, a rotation gate is employed to influence the excited state population. Conversely, in the absence of a spike, a ``delay'' gate, denoted by the symbol $\Delta$, is activated, allowing the qubit to remain inactive and naturally interact with the environment, thereby decaying to the ground state. After processing each input spike (or its absence), it is essential to measure the circuit to determine whether the excited state population has surpassed the firing threshold. If the threshold is surpassed, the state is reset to the ground state $\ket{0}$, and an output spike is registered. If it remains below the threshold, the subsequent spike must be processed.

Upon measuring the circuit, it must be reconstructed for each subsequent spike to be processed. This involves reconstructing the qubit's previous state by recreating the prior circuit while iterating with the latest spike's rotation or delay gate. This approach is inefficient and results in excessively deep circuits, which overlook the threshold mechanism. Fortunately, due to the use of only $R_X$ rotations and delay gates, the prior excited state population can be restored with a rotation gate set to the angle corresponding to the state before measurement. To formulate the time-discretized dynamics of the QLIF neuron akin to \Cref{eq:lif_discrete}, we examine the influence of each spike on the excited state population, $\alpha$. Initially, the `memory' effect must be treated as an action that restores the prior excited state population, allowing the subsequent input to be processed. To achieve this restoration, an $R_X$ gate should be applied using an angle of
\begin{equation}
    \label{eq:qlif_memory}
    \varphi[t] = 2\arcsin\left(\sqrt{\alpha[t]}\right).
\end{equation}

This formula describes the excited state population following a rotation applied to the quantum state. By inverting this relationship, we can recover the rotation angle that produced a given measured population. Although this may appear straightforward, it is a crucial step, particularly once the leaky (decay) dynamics are incorporated, as discussed in the following section.

For an input spike, corresponding to an $R_X(\theta)$ gate, leads to a measured state population of
\begin{equation}
    \label{eq:qlif_Theta}
    \Theta[t] = \sin^2\left(\frac{(\theta + \varphi[t])X[t]}{2}\right),
\end{equation}
for
\begin{equation}
\label{eq:X_spikes}
    X[t] = \begin{cases}
        1, \quad\text{spike,} \\
        0, \quad\text{no spike.}
    \end{cases}
\end{equation}
To represent the exponential decay due to noise and the $T_1$ relaxation process, we can simplify the treatment of open quantum systems and Markovian dynamics \cite{brand_markovian_2024}. In the basic scenario where an excited state relaxes to the ground state, the population decays exponentially with a characteristic time $T_1$, following $\propto \exp(-\tau/T_1)$ over a time interval $\tau$. On real quantum hardware, this effect occurs naturally due to inherent noise, and the parameter $\tau$ can be adjusted to control the decay rate. Within a quantum circuit, this relaxation is implemented using the ``Delay'' gate, $\Delta$, which causes the circuit to idle for a specified duration $\tau$, allowing the qubit to interact with its environment and decay toward the ground state.

For simulation and modeling purposes, this decay effect can be implemented manually. After the excited state population is increased by an $R_X$ gate, the exponential decay gradually reduces it over time. This decay can be replicated by applying a rotation in the opposite direction, with an angle chosen to match the expected decrease in population. The required angle is given by
\begin{equation}
    \label{eq:gamma_angle}
    \gamma[t] = -2\arcsin\left(\sqrt{\alpha[t] \exp\left(-\frac{\tau}{T_1}\right)}\right),
\end{equation}
after which it can take the same form of contribution to the excited state population as \Cref{eq:qlif_Theta}. Thus, as an expression for the dynamics of the excited state population, as a function of variables $\theta$ and $\tau$, we have
\begin{equation}
    \label{eq:qlif_main_cases}
    \alpha[t+1] = \begin{cases}
        \sin^2\left( \frac{\theta + \varphi[t]}{2} \right),\quad\,\,\,\,\,\text{if }X[t]=1, \\
        \sin^2\left( \frac{\gamma[t] + \varphi[t]}{2} \right),\quad\text{if }X[t]=0. \\
    \end{cases}
\end{equation}
This formulation expresses the dynamics as cases for the binary input, either a spike or no spike, mirroring the behavior of the quantum circuit during evaluation. The circuit reinstates the previous state (memory) and then applies either a positive rotation to represent a spike or a negative rotation to model the exponential decay (leakage) of the state. For the very first spike in the train, the memory reinstatement corresponds to a rotation of $0$, since the process always starts from the ground state. It is important to highlight that the circuit always initializes from the ground state, and the memory rotation encodes the neuron's prior activity. This approach ensures that only two gates are required per spike in the train, regardless of its length.

The excited state dynamics can also be written in a form that closely mirrors the classical expression in \Cref{eq:lif_final}. Rather than presenting the update as a set of cases, we can use the binary nature of $X[t]$ as a multiplicative selector, combining both cases into a single summation, just as in the classical model. Additionally, since $\sin^2(1) = \sin^2(-1)$, this allows for a compact formulation. Thus, the QLIF excited state population dynamics can be written as
\begin{equation}
    \label{eq:qlif_main}
    \begin{aligned}
    \alpha[t+1] &= \sin^2\left( \frac{\left(\theta + \varphi[t]\right) X[t]}{2} \right) \\
    &\quad+ \sin^2\left( \frac{\left(\gamma[t] + \varphi[t]\right)(X[t] - 1)}{2} \right).
    \end{aligned}
\end{equation}
In this form, the inputs and outputs can easily be vectorized, scaling as $\mathcal{O}(n)$, as compared to the simulation of other quantum circuits such as variational models, which scale as $\mathcal{O}(n^2)$ for $n$ qubits \cite{schuld2015introduction}. Furthermore, because the QLIF model only requires 2 gates and a single qubit for each evaluation, it eliminates all concern of noise disrupting the results, as is the case in deeper circuits of previous QML models.

\subsubsection{Spike Encoding}
\label{sec:Spike_Encoding}

Due to the versatility of these spiking neurons, there are also many ways to embed data within a framework that they can process. However, the fundamental link between all of these is the biological plausibility that they communicate and process based on spikes of current, or another transmitter, between neurons. In the case of biological LIF neurons, spikes of electrical impulses are transmitted along the synapses that connect them. In the quantum model, the spikes are treated more abstractly as a stimulus that surges excited state population.

To take a form of classical data, in whatever original form, and convert it to a spike train, there are a variety of options. The most common choices are rate coding, which uses Poisson statistics to generate a randomly distributed spike train with a spike frequency dependent on the input value as a spiking probability. Another option is temporal coding, which similarly converts the input value to a probability, but rather than a random distribution, the probability is linked to how early in the spike train the first spike will appear.

Although, as the focus of this work is centered on clustering algorithms, neither rate nor temporal coding can sufficiently translate the positional information to spikes in an effective and accurate way. With coordinate data, a greater distance from the origin would reflect a more important value, with a higher rate of spikes or earlier spikes, heavily influencing the neuronal dynamics in the wrong way. A solution to this is to map the coordinate grid to a grid of neurons that is aware of the rest of the grid.

This is done through \textit{population coding}. This encoding method uses multiple neurons to represent a continuous value or set of coordinates, distributing the representation across these neurons that each respond differently to the input, allowing for spatial orientation. Each neuron in the population has a preferred direction or value that defines it and a tuning curve that determines its response based on the similarity between the input and its preferred value. The response in this method typically comes in the form of rate coding, but now with an accurate connection between the input format and the spike train form \cite{eshraghian_training_2023}.

In context of processing coordinate data, the input vector is sent through a grid of receptive field neurons, corresponding to the preferred values, which each have a Gaussian tuning curve that responds strongest to an input closest to the receptive field. The grid of these receptive neurons is then collected and each spike train produced by these receptive fields is processed by a spiking neuron for a set of output spike trains. This is visualized in \Cref{fig:QLIF_SC}

\subsubsection{Van Rossum Distance Kernel}
\label{sec:van_Rossum_Distance_Kernel}

Once all of the coordinate points of a dataset have been processed through a spike encoding and spiking neuron pipeline, the spike trains need to be compared in some way to find patterns in the data. Classically, a kernel matrix is a distance metric between data points which can give a consolidated measure of similarity across all of the connections within the data, with each entry being a distance/similarity between one point and another. This concept can be easily extended to any data format, as long as a similarity metric can be defined. For spike trains, there are two leading methods to accomplish this: the van Rossum distance, and the Victor-Purpura distance.

The van Rossum (vR) distance is a flexible way of quantifying how different two sequences of action potentials are from one another \cite{vanRossum2001}. This is done by converting a train of discrete spikes into a continuous function by convolving each function with a kernel, such as an exponential decay function, of the form
\begin{equation}
    \label{eq:vr_kernel}
    f(t) = e^{-t/\tau}H(t),
\end{equation}
where $H(t)$ is the Heaviside step function, and $\tau$ is a time constant parameter which determines its sensitivity. For smaller to larger $\tau$ values, the vR distance behaves more like a coincidence detector to a rate difference counter, as its precision is changed through $\tau$.

After the spike trains have been converted to continuous functions, the Euclidean distance between them can be calculated as
\begin{equation}
    \label{eq:euclid_distance}
    D_\mathrm{vR}(X,Y) = \sqrt{\frac{1}{\tau} \int_{0}^{\infty} \left[ f_X(t) - f_Y(t) \right]^2 \dd{t}}
\end{equation}
where $f_X(t)$ and $f_Y(t)$ are the convolved spike trains.

\subsubsection{Victor-Purpura Distance Kernel}
\label{sec:Victor_Purpura_Distance_Kernel}

Another approach is the Victor-Purpura (VP) distance \cite{VictorPurpura1996,VictorPurpura1997}, which is often interchangeable with the vR distance depending on the context of the data. This method has a different approach to measuring spike train similarity, and does so through an \textit{edit distance}. It is a metric composed of the minimum cost to transform one spike train into another using only a set of elementary operations. These are: adding, deleting, or shifting a spike. Each of these operations has a tunable cost associated with it, to reward desired behavior in finding the path from one train to another.

For spike trains $X$ and $Y$, the VP distance is calculated as the sequence of operations that minimizes the total cost, as
\begin{equation}
    \label{eq:vp_distance}
    D_\mathrm{VP}(X,Y) = \min\left( \text{cost of transforming }X\text{ into }Y \right).
\end{equation}
This algorithm makes use of dynamic programming to efficiently find the minimum cost path. Similarly to the function of $\tau$ in the vR distance, the VP distance also has a time scale parameter, $q$, which controls the sensitivity of the algorithm to finding the optimal path. This parameter primarily influences the cost of shifting spikes, and ranges from a low value leading to the metric being a rate counter to a large value where the cost of shifting becomes more prohibitive and the algorithm behaves more as a coincidence detector.

Similarly to the output of the vR distance metric calculations, there are now values that define a similarity between each output spike train that represents each input coordinate of the dataset. These distances can be transformed into similarities through a Gaussian function as
\begin{equation}
    \label{eq:dist_to_kernel}
    K(i,j) = \exp\left( -\gamma D(i,j)^2 \right),
\end{equation}
where $D(i,j)$ is the distance (vR or VP) between the spike trains $i$ and $j$, and $\gamma$ is a scaling parameter that controls the decay rate. The resulting kernel matrix $K$ is symmetric and captures how similar the spike trains are in terms of their timing and structure. This kernel matrix can then be used in a multitude of ways, for any kernel-based methodology, but in the case of the present work it is used in the Spectral Clustering algorithm as a graph Laplacian. The complete process of going from coordinate data to spikes to a kernel matrix for clustering is summarized in \Cref{fig:QLIF_SC} for convenience.

\subsection{Evaluation Metrics and Majority Voting}

\begin{table}[b]
\captionsetup{justification=raggedright}
\centering
\renewcommand{\arraystretch}{1.4}
\begin{tabular}{p{0.3\linewidth} p{0.6\linewidth}}
\toprule
\textbf{Metric} & \textbf{Equation} \\
\midrule
Accuracy & $\frac{\text{TP} + \text{TN}}{\text{TP} + \text{FP} + \text{TN} + \text{FN}}$ \\
Precision & $\frac{\text{TP}}{\text{TP} + \text{FP}}$ \\
Recall & $\frac{\text{TP}}{\text{TP} + \text{FN}}$ \\
Silhouette Score & $\frac{b - a}{\max(a, b)}$ \\
Adjusted Rand Index (ARI) & $\frac{\text{RI} - \mathbb{E}[\text{RI}]}{\max(\text{RI}) - \mathbb{E}[\text{RI}]}$ \\
V-measure & $V = 2 \cdot \frac{h \cdot c}{h + c}$ \\
\bottomrule
\end{tabular}
\caption{Evaluation metrics for classification and clustering. TP, TN, FP, and FN denote true positives, true negatives, false positives, and false negatives, respectively. For clustering metrics: $a$ is intra-cluster distance, $b$ is nearest-cluster distance, RI is the Rand Index, and $h$, $c$ are homogeneity and completeness, respectively.}
\label{table:metrics_equations}
\end{table}

\begin{table*}[t!]
    \captionsetup{justification=raggedright}
    \centering
    \begin{tabular}{cccccc}
        \hline
        \multicolumn{6}{c}{\textbf{Classification Metrics}} \\
        \hline
        \textbf{Dataset} & \textbf{Regime} & \textbf{Kernel} & \textbf{Accuracy} & \textbf{Precision} & \textbf{Recall} \\
        \hline
        \multirow{2}{3em}{\texttt{Blobs}} 
        & Quantum & pQK & $0.91 \pm 0.086$ & $0.91 \pm 0.084$ & $0.91 \pm 0.084$ \\
        & Classical & RBF & $\mathbf{0.98 \pm 0.014}$ & $\mathbf{0.98 \pm 0.013}$ & $\mathbf{0.98 \pm 0.013}$ \\
        \hline
        \multirow{2}{3em}{\texttt{Circles}} 
        & Quantum & pQK & $0.85 \pm 0.024$ & $0.89 \pm 0.015$ & $0.86 \pm 0.021$ \\
        & Classical & RBF & $\mathbf{1.00 \pm 0.00}$ & $\mathbf{1.00 \pm 0.00}$ & $\mathbf{1.00 \pm 0.00}$ \\
        \hline
        \multirow{2}{3em}{\texttt{Moons}} 
        & Quantum & pQK & $0.79 \pm 0.087$ & $0.81 \pm 0.099$ & $0.78 \pm 0.086$ \\
        & Classical & RBF & $\mathbf{0.99 \pm 0.010}$ & $\mathbf{0.99 \pm 0.008}$ & $\mathbf{0.99 \pm 0.011}$ \\
        \hline
        \multirow{2}{3em}{\texttt{Iris}} 
        & Quantum & pQK & $0.85 \pm 0.040$ & $0.90 \pm 0.025$ & $0.85 \pm 0.036$ \\
        & Classical & RBF & $\mathbf{0.99 \pm 0.005}$ & $\mathbf{0.99 \pm 0.01}$ & $\mathbf{0.99 \pm 0.01}$ \\
        \hline
        \multirow{2}{3em}{\texttt{SDSS}} 
        & Quantum & pQK & $0.78 \pm 0.022$ & $0.57 \pm 0.028$ & $0.47 \pm 0.064$ \\
        & Classical & RBF & $\mathbf{0.81 \pm 0.037}$ & $\mathbf{0.60 \pm 0.01}$ & $\mathbf{0.52 \pm 0.07}$ \\
        \hline
        \multirow{2}{3em}{\texttt{SDSS}\textsuperscript{*}} 
        & Quantum & pQK & $0.59 \pm 0.048$ & $0.52 \pm 0.20$ & $\mathbf{0.57 \pm 0.059}$ \\
        & Classical & RBF & $\mathbf{0.56 \pm 0.13}$ & $\mathbf{0.68 \pm 0.13}$ & $0.56 \pm 0.13$ \\
        \hline
        \multicolumn{6}{c}{\textbf{Clustering Metrics}} \\
        \hline
        \textbf{Dataset} & \textbf{Regime} & \textbf{Kernel} & \textbf{Silhouette Score} & \textbf{ARI} & \textbf{V-measure} \\
        \hline
        \multirow{2}{3em}{\texttt{Blobs}} 
        & Quantum & pQK & $0.60 \pm 0.11$ & $0.80 \pm 0.16$ & $0.80 \pm 0.12$ \\
        & Classical & RBF & $\mathbf{0.64 \pm 0.073}$ & $\mathbf{0.95 \pm 0.041}$ & $\mathbf{0.93 \pm 0.058}$ \\
        \hline
        \multirow{2}{3em}{\texttt{Circles}} 
        & Quantum & pQK & $\mathbf{0.26 \pm 0.021}$ & $0.50 \pm 0.068$ & $0.52 \pm 0.040$ \\
        & Classical & RBF & $0.23 \pm 0.00$ & $\mathbf{1.00 \pm 0.00}$ & $\mathbf{1.00 \pm 0.00}$ \\
        \hline
        \multirow{2}{3em}{\texttt{Moons}} 
        & Quantum & pQK & $0.32 \pm 0.15$ & $0.34 \pm 0.17$ & $0.36 \pm 0.18$ \\
        & Classical & RBF & $\mathbf{0.40 \pm 0.029}$ & $\mathbf{0.97 \pm 0.039}$ & $\mathbf{0.94 \pm 0.054}$ \\
        \hline
        \multirow{2}{3em}{\texttt{Iris}} 
        & Quantum & pQK & $\mathbf{0.51 \pm 0.033}$ & $0.66 \pm 0.061$ & $0.76 \pm 0.030$ \\
        & Classical & RBF & $0.40 \pm 0.05$ & $\mathbf{0.95 \pm 0.04}$ & $\mathbf{0.95 \pm 0.04}$ \\
        \hline
        \multirow{2}{3em}{\texttt{SDSS}} 
        & Quantum & pQK & $0.35 \pm 0.019$ & $0.24 \pm 0.12$ & $0.23 \pm 0.080$ \\
        & Classical & RBF & $\mathbf{0.39 \pm 0.03}$ & $\mathbf{0.35 \pm 0.14}$ & $\mathbf{0.37 \pm 0.16}$ \\
        \hline
        \multirow{2}{3em}{\texttt{SDSS}\textsuperscript{*}} 
        & Quantum & pQK & $0.30 \pm 0.046$ & $0.18 \pm 0.013$ & $0.23 \pm 0.083$ \\
        & Classical & RBF & $\mathbf{0.36 \pm 0.04}$ & $\mathbf{0.20 \pm 0.15}$ & $\mathbf{0.27 \pm 0.16}$ \\
        \hline
    \end{tabular}
    \caption{Performance comparison between the parameterized quantum kernel and classical radial basis function kernel across synthetic and real datasets for classification and clustering. \texttt{SDSS}\textsuperscript{*} denotes the balanced version of the \texttt{SDSS} dataset. The best-performing result for each dataset is highlighted. For Silhouette Score, ARI, and V-measure, higher scores represent better clustering performance.}
    \label{tab:qkernel_and_classical_table}
\end{table*}

\begin{table*}[t!]
    \captionsetup{justification=raggedright}
    \centering
    \begin{tabular}{cccccc}
        \hline
        \multicolumn{6}{c}{\textbf{Classification Metrics}} \\
        \hline
        \textbf{Dataset} & \textbf{Neuron} & \textbf{Kernel} & \textbf{Accuracy} & \textbf{Precision} & \textbf{Recall} \\
        \hline
        \multirow{4}{3em}{\texttt{Blobs}} & \multirow{2}{3em}{LIF} & VP & \bm{$0.94 \pm 0.08$} & \bm{$0.94 \pm 0.08$} & \bm{$0.94 \pm 0.08$} \\
        && vR & $0.90 \pm 0.18$ & $0.90 \pm 0.21$ & $0.90 \pm 0.18$ \\
        & \multirow{2}{3em}{QLIF} & VP & $0.91 \pm 0.14$ & $0.91 \pm 0.13$ & $0.91 \pm 0.14$ \\
        && vR & $0.88 \pm 0.15$ & $0.86 \pm 0.20$ & $0.88 \pm 0.15$ \\
        \hline
        \multirow{4}{3em}{\texttt{Circles}} & \multirow{2}{3em}{LIF} & VP & \bm{$0.99 \pm 0.004$} & \bm{$0.99 \pm 0.004$} & \bm{$0.99 \pm 0.004$} \\
        && vR & $0.59 \pm 0.07$ & $0.78 \pm 0.02$ & $0.59 \pm 0.07$ \\
        & \multirow{2}{3em}{QLIF} & VP & $0.99 \pm 0.01$ & $0.99 \pm 0.01$ & $0.99 \pm 0.01$ \\
        && vR & $0.64 \pm 0.10$ & $0.79 \pm 0.04$ & $0.64 \pm 0.10$ \\
        \hline
        \multirow{4}{3em}{\texttt{Moons}} & \multirow{2}{3em}{LIF} & VP & $0.85 \pm 0.02$ & $0.86 \pm 0.01$ & $0.85 \pm 0.02$ \\
        && vR & $0.81 \pm 0.13$ & $0.86 \pm 0.06$ & $0.81 \pm 0.13$ \\
        & \multirow{2}{3em}{QLIF} & VP & \bm{$0.87 \pm 0.04$} & \bm{$0.88 \pm 0.02$} & \bm{$0.87 \pm 0.04$} \\
        && vR & $0.90 \pm 0.11$ & $0.93 \pm 0.06$ & $0.90 \pm 0.11$ \\
        \hline
        \multirow{4}{3em}{\texttt{Iris}} & \multirow{2}{3em}{LIF} & VP & $0.88 \pm 0.04$ & $0.91 \pm 0.02$ & $0.88 \pm 0.04$ \\
        && vR & $0.88 \pm 0.04$ & $0.91 \pm 0.02$ & $0.88 \pm 0.04$ \\
        & \multirow{2}{3em}{QLIF} & VP & \bm{$0.89 \pm 0.02$} & \bm{$0.91 \pm 0.01$} & \bm{$0.89 \pm 0.02$} \\
        && vR & $0.90 \pm 0.02$ & $0.91 \pm 0.02$ & $0.90 \pm 0.02$ \\
        \hline
        \multirow{2}{3em}{\texttt{SDSS}} & LIF & VP & $0.71 \pm 0.002$ & $0.27 \pm 0.11$ & $0.33 \pm 0.004$ \\
        & QLIF & VP & \bm{$0.71 \pm 0.001$} & \bm{$0.27 \pm 0.11$} & \bm{$0.33 \pm 0.002$} \\
        \hline
        \multirow{2}{3em}{\texttt{SDSS}\textsuperscript{*}} & LIF & VP & \bm{$0.43 \pm 0.06$} & \bm{$0.69 \pm 0.14$} & \bm{$0.43 \pm 0.06$} \\
        & QLIF & VP & $0.40 \pm 0.07$ & $0.62 \pm 0.20$ & $0.40 \pm 0.07$ \\
        \hline
        \multicolumn{6}{c}{\textbf{Clustering Metrics}} \\
        \hline
        \textbf{Dataset} & \textbf{Neuron} & \textbf{Kernel} & \textbf{Silhouette Score} & \textbf{ARI} & \textbf{V-measure} \\
        \hline
        \multirow{4}{3em}{\texttt{Blobs}} & \multirow{2}{3em}{LIF} & VP & \bm{$0.60 \pm 0.15$} & \bm{$0.87 \pm 0.16$} & \bm{$0.86 \pm 0.15$} \\
        && vR & $0.56 \pm 0.27$ & $0.82 \pm 0.29$ & $0.82 \pm 0.27$ \\
        & \multirow{2}{3em}{QLIF} & VP & $0.62 \pm 0.16$ & $0.82 \pm 0.24$ & $0.82 \pm 0.22$ \\
        && vR & $0.61 \pm 0.14$ & $0.81 \pm 0.22$ & $0.85 \pm 0.16$ \\
        \hline
        \multirow{4}{3em}{\texttt{Circles}} & \multirow{2}{3em}{LIF} & VP & \bm{$0.23 \pm 0.01$} & \bm{$0.96 \pm 0.02$} & \bm{$0.93 \pm 0.02$} \\
        && vR & $0.27 \pm 0.07$ & $0.05 \pm 0.09$ & $0.13 \pm 0.09$ \\
        & \multirow{2}{3em}{QLIF} & VP & $0.23 \pm 0.01$ & $0.96 \pm 0.03$ & $0.93 \pm 0.04$ \\
        && vR & $0.33 \pm 0.04$ & $0.11 \pm 0.17$ & $0.20 \pm 0.15$ \\
        \hline
        \multirow{4}{3em}{\texttt{Moons}} & \multirow{2}{3em}{LIF} & VP & $0.45 \pm 0.03$ & $0.49 \pm 0.06$ & $0.43 \pm 0.04$ \\
        && vR & $0.36 \pm 0.19$ & $0.44 \pm 0.25$ & $0.43 \pm 0.18$ \\
        & \multirow{2}{3em}{QLIF} & VP & \bm{$0.47 \pm 0.04$} & \bm{$0.55 \pm 0.10$} & \bm{$0.47 \pm 0.07$} \\
        && vR & $0.39 \pm 0.17$ & $0.70 \pm 0.26$ & $0.66 \pm 0.22$ \\
        \hline
        \multirow{4}{3em}{\texttt{Iris}} & \multirow{2}{3em}{LIF} & VP & $0.42 \pm 0.02$ & $0.69 \pm 0.07$ & $0.73 \pm 0.04$ \\
        && vR & $0.40 \pm 0.03$ & $0.70 \pm 0.06$ & $0.73 \pm 0.03$ \\
        & \multirow{2}{3em}{QLIF} & VP & \bm{$0.45 \pm 0.01$} & \bm{$0.72 \pm 0.04$} & \bm{$0.74 \pm 0.02$} \\
        && vR & $0.44 \pm 0.01$ & $0.72 \pm 0.04$ & $0.73 \pm 0.03$ \\
        \hline
        \multirow{2}{3em}{\texttt{SDSS}} & LIF & VP & $-0.16 \pm 0.09$ & $0.002 \pm 0.007$ & $0.003 \pm 0.009$ \\
        & QLIF & VP & \bm{$-0.13 \pm 0.07$} & \bm{$0.001 \pm 0.004$} & \bm{$0.001 \pm 0.005$} \\
        \hline
        \multirow{2}{3em}{\texttt{SDSS}\textsuperscript{*}} & LIF & VP & \bm{$-0.16 \pm 0.10$} & \bm{$0.03 \pm 0.03$} & \bm{$0.12 \pm 0.06$} \\
        & QLIF & VP & $-0.19 \pm 0.10$ & $0.03 \pm 0.05$ & $0.08 \pm 0.08$ \\
        \hline
    \end{tabular}
    \caption{Neuromorphic classification and clustering results on all datasets. \texttt{SDSS}\textsuperscript{*} denotes a balanced sampling of the \texttt{SDSS} dataset. The best-performing result for each dataset is highlighted. For Silhouette Score, ARI, and V-measure, higher scores represent better clustering performance.}
    \label{tab:neuro_full_results_synth}
\end{table*}

In order to compare clustering performance using different origins for the kernel matrix, we adopted three evaluation strategies. First, we used clustering metrics, Silhouette Score, Adjusted Rand Index (ARI), and V-measure, to assess overall clustering quality. For each dataset, we plotted these metrics against a range of cluster counts \( K \), producing an ``elbow-like" curve inspired by the elbow method in \( K \)-means clustering \cite{alpaydin2020introduction, murphy2012machine}. We will refer to these curves as clustering metric peak plots from now on. The value of \( K \) at which these metrics reach their maximum was taken as an indication of the optimal number of clusters. Finally, to enable comparison with ground truth labels, we performed a label mapping via majority voting within clusters and computed classification metrics, accuracy, precision, and recall, on the reassigned labels. All metrics are calculated using built-in functions in \texttt{scikit-learn}, but their definitions can be found in Table \ref{table:metrics_equations}.

True Positives (TP) refer to the number of samples that are correctly identified as belonging to a given class. True Negatives (TN) count the samples correctly identified as not belonging to that class. False Positives (FP) are those incorrectly classified as belonging to the class, while False Negatives (FN) are samples that should have been classified as such but were missed.

The intra-cluster distance, denoted by $a$, is the average distance between points within the same cluster. It reflects the internal compactness or similarity of a cluster. The nearest-cluster distance, $b$, is the average distance from each point to points in the closest neighboring cluster and captures how well-separated the clusters are from each other.

The Rand Index (RI) is used to compare two clusterings by measuring the proportion of sample pairs that are either grouped together or separated in both clusterings:
\begin{equation}
\text{RI} = \frac{\text{number of agreements}}{\text{total pairs}}.
\end{equation}

Homogeneity, denoted by $h$, assesses whether each cluster contains only members of a single class. It is defined as:
\begin{equation}
h = 1 - \frac{H(C|K)}{H(C)},
\end{equation}
where $H(C|K)$ is the conditional entropy of the true class labels given the predicted clusters, and $H(C)$ is the entropy of the true labels.

Completeness, written as $c$, evaluates whether all samples of a given class are assigned to the same cluster. It is calculated as:
\begin{equation}
c = 1- \frac{H(K|C)}{H(K)},
\end{equation}
where $H(K|C)$ is the conditional entropy of the clustering given the true class labels, and $H(K)$ is the entropy of the clustering. Both homogeneity and completeness range from 0 to 1, with 1 indicating perfect homogeneity or completeness.

Accuracy measures the proportion of correctly predicted samples out of all predictions made. Precision refers to the proportion of true positive predictions among all predicted positives, while recall is the proportion of true positive predictions among all actual positive samples. All of these classification metrics range from 0 to 1. The Silhouette Score quantifies how well each sample fits within its cluster compared to other clusters, ranging from -1 to 1. The Adjusted Rand Index (ARI) compares the similarity between the predicted and true clusterings while adjusting for chance, ranging from -1 (poor) to 1 (perfect). The V-measure is the harmonic mean of homogeneity and completeness, capturing how well clusters reflect the actual class structure, and ranges from 0 to 1.

In order to connect spectral clustering, which is inherently unsupervised, to evaluation metrics that rely on ground-truth labels, a label mapping strategy must be employed. After clustering, the resulting cluster labels are arbitrary and do not correspond directly to the true class labels. To facilitate meaningful evaluation, we apply a majority voting scheme \cite{alpaydin2020introduction, murphy2012machine}: for each cluster, we identify the most frequent class label among the data points assigned to that cluster and assign this label to all points within the cluster. This procedure aligns the arbitrary cluster indices with the actual class labels as closely as possible. Once this mapping is established, the clustering performance can be quantitatively assessed using supervised metrics such as accuracy, precision, and recall.

\section{Results}
\label{sec:Results}

\begin{figure*}[t!]
    \centering
    \includegraphics[width=\textwidth]{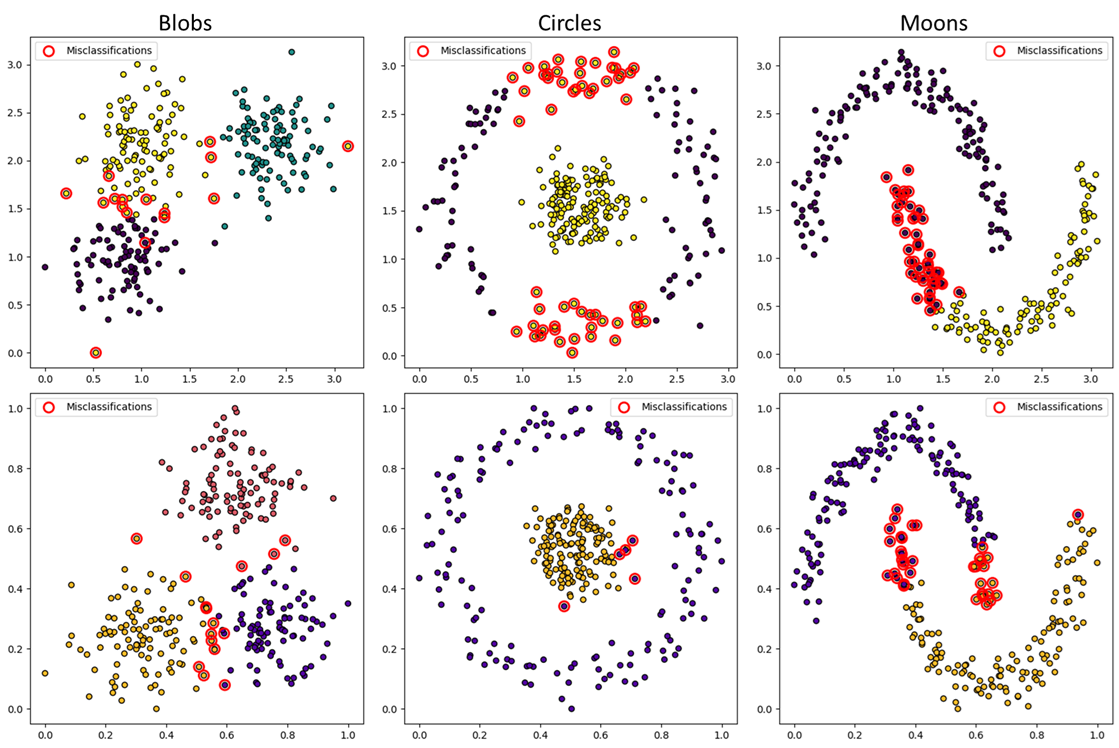}
    \captionsetup{justification=raggedright}
    \caption{Visualization of label predictions through majority voting after clustering. Red circles indicate samples that are predicted incorrectly. Top: pQK, Bottom: QLIF VP.}
    \label{fig:spectral_plots}
\end{figure*}

\begin{figure*}[t!]
    \centering
    \includegraphics[width=\textwidth]{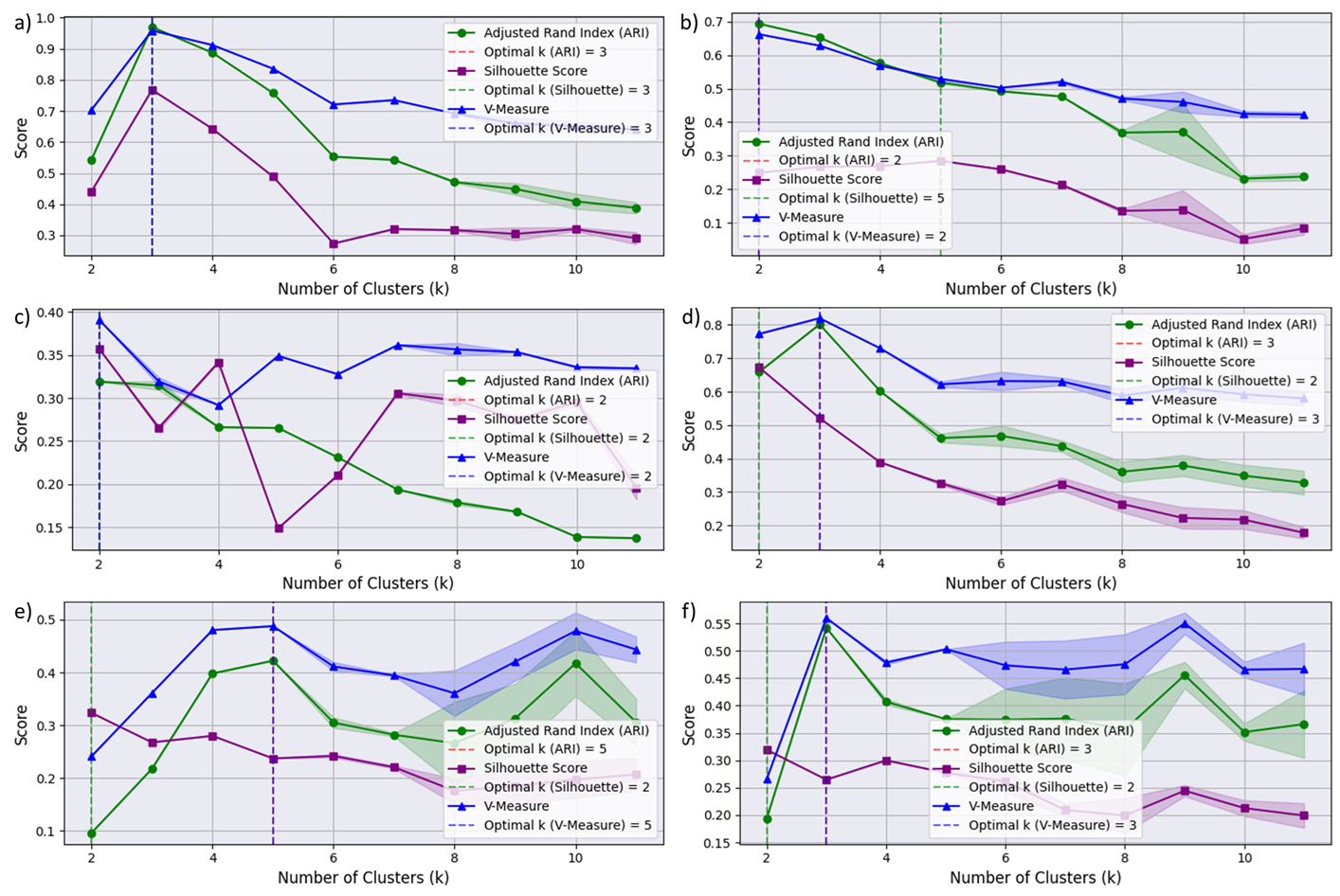}
    \captionsetup{justification=raggedright}
    \caption{Clustering metric peak plots for each dataset using the parameterized quantum kernel during clustering. Datasets indicated: \textbf{a)} \texttt{Blobs}, \textbf{b)} \texttt{Circles}, \textbf{c)} \texttt{Moons}, \textbf{d)} \texttt{Iris}, \textbf{e)} \texttt{SDSS}, \textbf{f)} \texttt{SDSS}\textsuperscript{*}. In each frame the ARI, Silhouette Score, and V-Measure over a varying number of clusters are plotted. The highest of each metric indicates the ideal number of clusters for that metric. The expected cluster number for each dataset is: \textbf{a)} 3, \textbf{b)} 2, \textbf{c)} 2, \textbf{d)} 3, \textbf{e)} 3, \textbf{f)} 3.}
    \label{fig:trainable_elbows}
\end{figure*}

\begin{figure*}[t!]
    \centering
    \includegraphics[width=\textwidth]{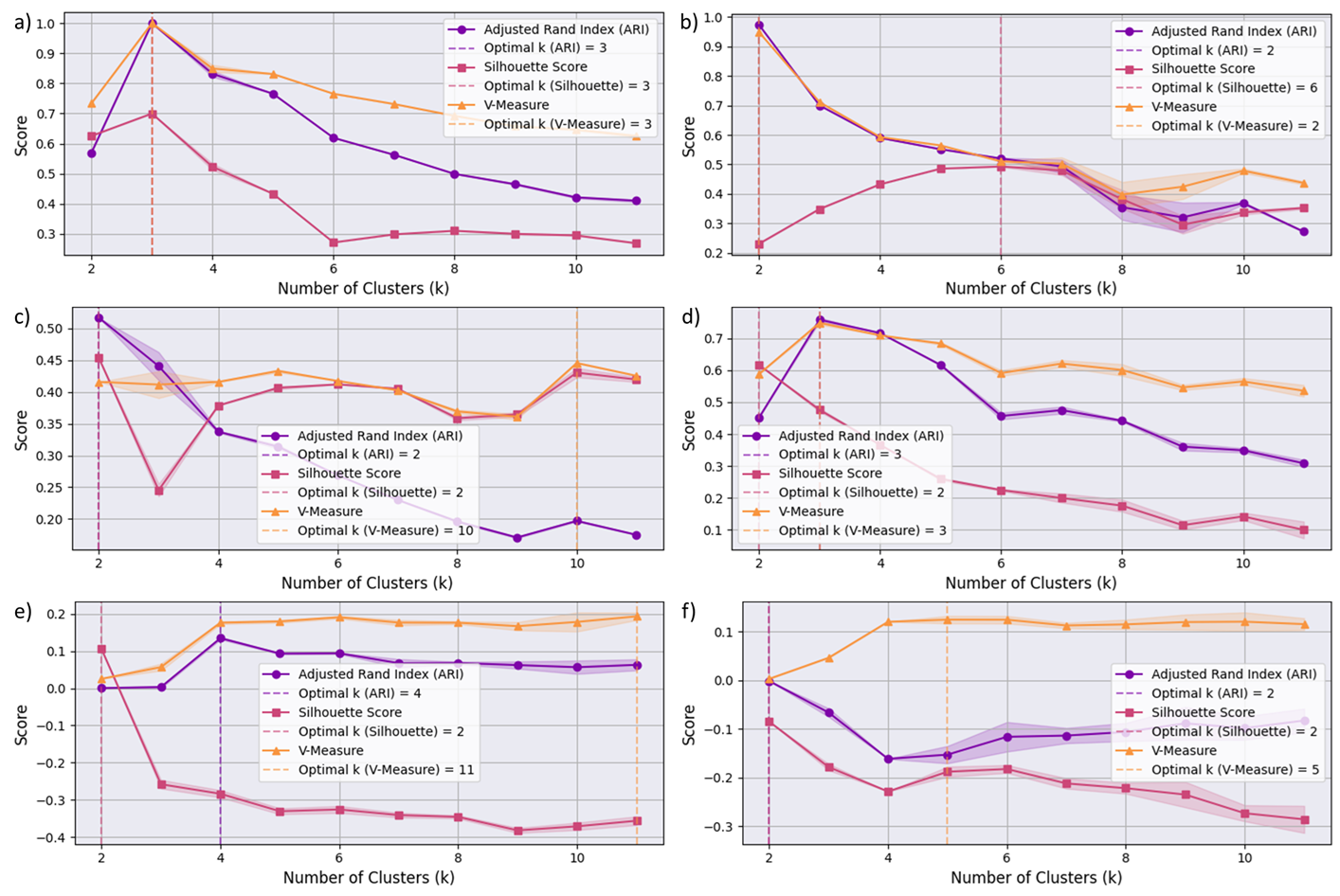}
    \captionsetup{justification=raggedright}
    \caption{Clustering metric peak plots for each dataset using the QLIF neuron in the neuromorphic clustering algorithm. Datasets indicated: \textbf{a)} \texttt{Blobs}, \textbf{b)} \texttt{Circles}, \textbf{c)} \texttt{Moons}, \textbf{d)} \texttt{Iris}, \textbf{e)} \texttt{SDSS}, \textbf{f)} \texttt{SDSS}\textsuperscript{*}. In each frame the ARI, Silhouette Score, and V-Measure over a varying number of clusters are plotted. The highest of each metric indicates the ideal number of clusters for that metric. The expected cluster number for each dataset is: \textbf{a)} 3, \textbf{b)} 2, \textbf{c)} 2, \textbf{d)} 3, \textbf{e)} 3, \textbf{f)} 3.}
    \label{fig:qlif_elbows}
\end{figure*}

\subsection{Comparison: Parameterized Quantum Kernel and Classical RBF Kernel}

In \Cref{tab:qkernel_and_classical_table}, it is clear that the classical RBF kernel, with $\gamma = 10$, consistently outperforms the parameterized quantum kernel in nearly all cases. This result is expected, as the RBF kernel is a well-established, highly expressive kernel that performs particularly well on smooth and well-separated datasets, such as the synthetic ones considered here. In fact, it achieves near-perfect classification metrics on datasets like \texttt{Blobs}, \texttt{Circles}, and \texttt{Moons}. In contrast, our quantum kernel is intentionally kept simple, using only three layers of parameterized rotation gates. Although this simplicity helps with interpretability, limits gate-based noise, and reduces resource requirements, it also limits the expressivity of the quantum kernel compared to its classical counterpart.

Despite the inferiority of the metrics for the quantum kernel, the results are generally quite competitive. The classification metrics for \texttt{Iris} and the synthetic datasets never drop below 0.75, except for the \texttt{SDSS} results, although even the classical RBF kernel struggled in that case. In particular, while RBF maintains higher precision and accuracy in real-world datasets such as \texttt{Iris} and \texttt{SDSS}, the quantum kernel achieves relatively strong recall on the balanced \texttt{SDSS}\textsuperscript{*} dataset, though it still trails in overall performance. The clustering metrics tell a different story. Although the silhouette scores are comparable across the board, the pQK performed markedly worse in some cases when considering ARI and V-measure. Note the values for \texttt{Circles}, \texttt{Moons}, and \texttt{Iris}.

\subsection{Comparison: QLIF and LIF}

In the neuromorphic spiking neuron kernels, there is a much greater match between the quantum and classical cases, as can be seen in \Cref{tab:neuro_full_results_synth}. This is expected because the pipelines are very similar due to the design of the quantum LIF neuron as a direct replacement for the classical model. These slight differences in the behavior of the neurons are not, however, a disadvantage as may be expected from a model that does not utilize the typical strengths of qubits: superposition and entanglement. The simple quantum circuit comprised of a single qubit and two rotation gates performs very closely to the classical model or outperforms it in some cases.

In the scenario of the \texttt{Blobs} dataset, there is a slight loss in classification performance, but more variability, which is less reliable on average but potentially performs better with certain distributions. In the circles dataset both models perform identically, and near perfectly. This demonstrates the ability of the spiking kernel construction as a versatile and effective algorithm for clustering non-linear data, compared to $K$-means, for example, which cannot capture the clusters in this manner.

However, it is also clear from this example especially that the choice of distance metric, between the van Rossum and Victor-Purpura methods, has a significant impact on this methodology. While they may be comparable and interchangeable in other cases, with their own relative advantages dependent on the application, in the case of this clustering pipeline there is a clear preferable choice. The VP metric leads to consistently higher classification and clustering results. VP runs were typically faster to calculate. For this reason, it is the only kernel evaluated for the \texttt{SDSS} datasets, which already take significantly longer than the synthetic datasets, and would only take longer with the vR-based kernel with predictably worse results.

What is also clear from the results in the more complicated datasets, in terms of non-linearity such as \texttt{Moons} and \texttt{Iris}, is the advantage gained by using the QLIF neuron over its classical counterpart. Albeit a small advantage, it is significant enough to make it a preferable option to the classical method, which is not common for quantum models because of the drawbacks often associated with quantum algorithms applied to classical data, such as circuit complexity, calculation time, and data encoding.

Unfortunately, all of the neuromorphic models struggle to accurately and consistently cluster and classify the \texttt{SDSS} datasets, and take significantly longer than non-spiking methods. This is largely due to the complexity involved in expanding the population coding scheme to account for the higher dimensionality of the dataset, which scales exponentially. Due to this scaling, there is a harsh penalty to computation time, which is traded off for lower resolution in the encoding. This is also seen in the clustering metrics, which show that the model struggles to resolve any of the clusters, with the classification being random guessing at best.

\subsection{Comparison: pQK versus QLIF}

This section directly compares the pQK and quantum neuromorphic kernels. Since the VP kernel with QLIF seems to be the more promising of the two, all the focus in this comparison will be placed on it.

When comparing the parameterized quantum kernel (pQK) with the QLIF Victor-Purpura kernel, QLIF consistently performs better in classification for the synthetic data and for \texttt{Iris}. For instance, on the \texttt{Blobs} dataset, QLIF VP achieves 0.94 across all classification metrics, surpassing pQK’s 0.91. This trend persists in \texttt{Circles} and \texttt{Moons}, where QLIF VP provides better precision and recall. These results suggest that QLIF VP captures non-linearities and class boundaries better than pQK, offering a strong case for its use as a quantum neuromorphic kernel for spectral clustering. It is important to note that this is true in the comparison against a general three-layer, unentangled quantum kernel. The expectation is that without applying parameterized scaling to the data encoding rotations, the pQK would have performed worse compared to QLIF. For the \texttt{SDSS} cases, pQK performed better overall. 

For comparison using clustering metrics the trend is that QLIF VP once again edges out over pQK in the all three of the synthetic datasets, however, for \texttt{Iris} it depends on which metric we consider. pQK is superior in terms of Silhouette Score and V-measure. QLIF VP fails for both \texttt{SDSS} and \texttt{SDSS}$\*$. pQK performs markedly better as in the classification metrics, but it is important to note that in terms of what is considered good clustering metrics, the results can still be considered suboptimal. 

The intuition for this could be the trainability of the angle encoding scheme. Input features are scaled by parameters before being mapped to quantum rotation gates, allowing the feature map to adapt during training and align the quantum kernel more closely with the underlying data structure. Rather than relying on fixed projections, the model learns how to encode features in a way that supports effective spectral clustering. Although these parameters were constrained to positive values (e.g., $[0, 2]$), extending the range to include negative values (e.g., $[–2, 2]$) could further enhance the expressive power by capturing both the magnitude and direction of feature differences, potentially revealing hidden sign-based correlations. QLIF converts samples into spike trains before being input into the neuromorphic kernels. The primary struggle of the spiking methodology in the spectral clustering lies with the population coding of the data. Due to this encoding scheme's reliance on the Gaussian distancing between preferred data points, there needs to be a definition of nodes in the grid space that define the distances and the effective resolution of the data space. However, this then scales exponentially with dimension, both in time and in computational complexity, leading to limitations in how highly resolved the nodes can be. This limitation leads not only to a longer computation time, but also to less accurate classification and clustering. Due to the trainability of pQK, and since both the \texttt{Iris} and \texttt{SDSS} datasets are overlapping and high-dimensional (unlike the 2D \texttt{Moons} and \texttt{Circles} datasets, which are non-overlapping but not linearly separable) this could explain the superior performance of pQK for \texttt{SDSS}. 

\Cref{fig:spectral_plots} is a visualization of the clustering performance for an example set generated for each of the synthetic datasets. Misclassifications, indicated in red circles, can be observed close to overlapping regions for \texttt{Blobs}. For \texttt{Circles} and \texttt{Moons} the difference in performance can be seen in how only some central samples are misclassified when applying QLIF to \texttt{Circles} and how pQK managed to correctly predict all samples from one class in \texttt{Moons}, but perform much worse for the other class.

\subsection{Optimal number of clusters k}

Using the clustering metric peak plots from \Cref{fig:trainable_elbows} and \Cref{fig:qlif_elbows} as elbow-like curves for spectral clustering we are able to find the optimal number of clusters to use per dataset. The hope is that the optimal number of clusters would correspond to the total number of classes in the dataset. We will consider this optimal number of clusters output to be a prediction and call it correct if the optimal number of clusters and the true number of classes are equal. This may not always be the case, as real-world data often contains complex or overlapping structures where, for instance, three classes may yield an optimal clustering at five clusters. To address this ambiguity, we include synthetic datasets where the relationship between classes and clusters is more direct and well defined.

For the QLIF VP kernel, the predicted number of clusters matched the true number of classes for the \texttt{Blobs} dataset, correctly identifying three clusters. In the \texttt{Circles} dataset, the clustering metrics peaked at two and six clusters, with ARI and V-measure supporting the correct value. Similarly, for the \texttt{Moons} dataset, peaks occurred at two and ten clusters, with ARI and Silhouette Score being correct. For the \texttt{Iris} dataset, the curves suggested two and three clusters, where ARI and V-measure correctly aligned with expectations. This is a reasonable result given the known overlap between two of the classes in the \texttt{Iris} dataset. In contrast, the \texttt{SDSS} dataset yielded peaks at four, two, and eleven clusters, none of which matched the actual number of classes. The same trend was observed in \texttt{SDSS}\textsuperscript{*}, where peaks at two and five clusters did not correspond to the correct class count.

For the pQK kernel, the predicted cluster count also aligned well with the \texttt{Blobs} dataset, correctly suggesting three clusters. In the \texttt{Circles} dataset, predictions peaked at two and five clusters, with ARI and V-measure identifying the correct number. The \texttt{Moons} dataset consistently produced a correct prediction of two clusters. As with QLIF VP, the \texttt{Iris} dataset showed peaks at two and three clusters, both supported by ARI and V-measure. For the \texttt{SDSS} dataset, predictions of two and five clusters did not match the true class count. However, in \texttt{SDSS}\textsuperscript{*} variant, peaks at two and three clusters included a correct prediction, where ARI and V-measure were correct.

The results suggest the use of either V-measure or ARI for finding the optimal number of clusters to use in this elbow-like approach for spectral clustering. The results are also consistent with the previous observation that pQK fairs better with \texttt{SDSS}.

\section{Conclusion}
\label{sec:Conclusion}

We directly compared a parameterized quantum kernel with two quantum neuromorphic kernels, the Victor-Purpura and van Rossum kernels. We also included the RBF kernel to serve as a classical benchmark. For the pQK case two unitary gates are required: one unitary that encodes one feature vector and a second unitary that encodes a second feature vector. The second unitary is the adjoint of the first. A projector matrix allows for the calculation of a fidelity value without the use of a SWAP gate. The encoding is done through three layers in each unitary, one for each direction in the Bloch sphere. These rotations are also scaled by parameters. An optimal set of parameters are found through a grid search in the hope of finding an angle encoded quantum kernel that is representative of distance in feature space. This is done by maximizing the kernel target alignment. The neuromorphic case relies on population coding to transform samples into spike trains, which are then ultimately passed to the two kernels. Once the kernel matrices were calculated, we applied classical spectral clustering. The overall performance of the pipelines was compared using label-based classification and clustering metrics, as well as finding the optimal number of clusters in an elbow-like approach.

In general, QLIF is typically observed to dominate with respect to classification and clustering metrics for most datasets. Only during the higher-dimensional \texttt{SDSS} did pQK perform better. The explanation for this is a combination of the trainability of pQK and the limitations of QLIF at higher dimensions. This result is consistent with the optimal number of clusters predicted by the elbow-like plots in \Cref{fig:trainable_elbows} and \Cref{fig:qlif_elbows}. pQK correctly predicted the optimal number k for balanced \texttt{SDSS}. Please note that we considered applying PCA before the clustering pipeline in order to remove the pQK's dimensionality advantage by dimensionality reduction to two dimensions. This would also have made visualization possible; however, all results were markedly worse. Since the \texttt{SDSS} dataset used is already a feature-selected subset of the original dataset, we decided not to include these results. We also considered applying PCA or t-SNE after clustering purely for visualization purposes. We ultimately decided against inclusion, as it would not be a true representation of clustering in feature space, only a projection, which did not contribute new information. We believe a better alternative would be to simply feature select two features at a time and visualize clustering in a pairwise basis.

The pQK could be improved by increasing the capacity or complexity of the quantum model. This could potentially be achieved by including additional sets of encoding gates or by incorporating entangling gates, specifically controlled rotation gates. Another way to improve this would be to allow the scaling parameters to take values from $[-2, 2]$ instead of just $[0, 2]$. In this way, we also allow for negative rotations. In either case, the entire Bloch sphere is probed. In the neuromorphic case, there is a wide array of options to explore for the modular pipeline, ranging from different encoding schemes to distance metrics. Investigating performance comparisons with these changes is an interesting avenue of improvement. Finally, another way to expand on this comparison, is to eventually run the full pipeline on quantum hardware. This is currently impossible because running a number of quantum circuit executions that are quadratically scaling is highly infeasible.

\begin{acknowledgments}
We wish to acknowledge the International Year of Quantum Science and Technology. We would also like to thank our colleagues, Matt Lourens and Amy Rouillard, for their input during early discussions.
\end{acknowledgments}

\section*{Funding}

This work was funded by the South African Quantum Technology Initiative (SA QuTI) through the Department of Science, Technology and Innovation of South Africa. This work is based on research supported in part by the National Research Foundation of South Africa, Ref. PMDS22070532362. The funders played no role in the study design, data collection, analysis and interpretation of data, or the writing of this manuscript.

\section*{Data Availability Statement}
\label{sec:data_availability}

All synthetic datasets (\texttt{Blobs}, \texttt{Circles}, and \texttt{Moons}) are generated using built-in \texttt{scikit-learn} functions. The \texttt{Iris} dataset can be imported from \texttt{sklearn.datasets}, and the \href{https://zenodo.org/records/3768398}{\texttt{SDSS}} is available for download \cite{clarke2020sdss}.

\section*{Conflict of Interest}
The authors declare no conflict of interest.

\bibliography{references.bib}

\end{document}